\documentclass[twocolumn]{openjournal}
\usepackage{graphicx} 
\usepackage{multirow}

\usepackage[colorlinks,linkcolor=blue,citecolor=blue,urlcolor=blue ]{hyperref}
\usepackage[utf8]{inputenc}
\usepackage{float}
\usepackage{xcolor}
\usepackage{ulem}
\usepackage[T1]{fontenc}
\usepackage{amsmath}
\usepackage{xspace}
\usepackage{array,booktabs} 
\newcolumntype{P}[1]{>{\raggedright\arraybackslash}p{#1}} 

\newcommand{\msolar}{$\rm{M_{\odot}~}$}
\newcommand{\msolarc}{$\rm{M_{\odot}}$}
\newcommand{\zsolar}{$\rm{Z_{\odot}~}$}

\defcitealias{Prole2023}{LP23}
\defcitealias{Schauer2021}{AS21}

\defcitealias{Prole2026}{LP26}
\begin{document}


\title{SEEDZ: Rapid Galaxy Assembly as a Pathway to Supermassive Stars, Dense Stellar Environments  and Massive Black Hole Seeds \vspace{-1cm}}
\author{Lewis R. Prole$^{* \ 1}$}
\author{John A. Regan$^{1}$}
\author{Daxal Mehta$^{1}$}
\author{Devesh Nandal$^{2}$}
\author{R\"udiger Pakmor$^{3}$}
\author{Ricarda S. Beckmann$^{4}$}
\author{Michael Tremmel$^{5}$}
\author{Martin G. Haehnelt$^{6}$}
\author{Simon C.~O. Glover$^{7}$}
\author{Ralf S. Klessen$^{7}$}
\author{John H. Wise$^{8}$}
\author{Sophie Koudmani$^{9}$}
\author{Martin A. Bourne$^{6,9}$}
\author{Debora Sijacki$^{6}$}
\author{John Brennan$^{1}$}
\author{Pelle van de Bor$^{1}$}
\author{Paul C. Clark$^{10}$}
\email{$^*$email: lewis.prole@mu.ie}
\affiliation{\\ \ \ \ \ $^{1}$ Centre for Astrophysics and Space Sciences Maynooth, Department of Physics, Maynooth University, Maynooth, Ireland.}
\affiliation{$^{2}$ Center for Astrophysics, Harvard and Smithsonian, 60 Garden St, Cambridge, MA 02138, USA}
\affiliation{$^{3}$ Max-Planck-Institut f\"ur Astrophysik, Karl-Schwarzschild-Straße 1, D-85748, Garching, Germany}
\affiliation{$^{4}$ Institute for Astronomy, University of Edinburgh, Royal Observatory, Edinburgh EH9 3HJ, UK}
\affiliation {$^{5}$ School of Physics, University College Cork, College Road, Cork T12 K8AF, Ireland}
\affiliation {$^{6}$ Kavli Institute of Cosmology, Cambridge, University of Cambridge, Madingley Road, Cambridge CB3 0HA, UK}
\affiliation {$^{7}$ Universit\"{a}t Heidelberg, Zentrum f\"{u}r Astronomie, Institut f\"{u}r Theoretische Astrophysik, Albert-Ueberle-Stra{\ss}e 2, 69120 Heidelberg, Germany.}
\affiliation {$^{8}$ Center for Relativistic Astrophysics, School of Physics, Georgia Institute of Technology, 837 State Street, Atlanta, GA 30332, USA}
\affiliation{$^{9}$ Centre for Astrophysics Research, Department of Physics, Astronomy and Mathematics, University of Hertfordshire, College Lane, Hatfield, AL10 9AB, UK}
\affiliation {$^{10}$ Cardiff University School of Physics and Astronomy}


\begin{abstract}
\noindent 
  We investigate the assembly history of early galaxies in the \texttt{SEEDZ} hydrodynamic simulations, to investigate the high inflow rates believed to be required for the formation of supermassive stars (SMSs), dense stellar clusters and subsequently heavy seed black holes. Using a heavy seed formation criteria of $>$1 M$_\odot$ yr$^{-1}$ flowing into 10 pc regions, we find that heavy seeds form in halos that grow rapidly compared to those halos that never meet the criteria. Halos with growth rates of $\gtrsim$1 M$_\odot$ yr$^{-1}$ at their virial radius (scales of a few hundred pc) are able to sustain a flow rate of 0.1 M$_\odot$ yr$^{-1}$ into the inner 1 pc of the halo, maintaining higher density environments within the central 10 - 100~pc. These halos continue to grow rapidly after their initial collapse, typically forming heavy seeds $\sim$100 Myr after forming their first stars and stellar mass black holes. By $z=10$, most heavy seeds form in regions of near-solar metallicity, although a minority of heavy seeds do continue to form in low metallicity (10$^{-2}$ Z$_\odot$) regions. Under the assumption that a SMS forms as the progenitor to a heavy seed if it forms in a region of low (10$^{-2}$ Z$_\odot$) metallicity, and can sustain high accretion rates above 0.02 M$_\odot$ yr$^{-1}$ throughout the SMS lifetime of 2 Myr, we find a number density of SMSs of 0.1 cMpc$^{-3}$, meaning that only a fraction of 10$^{-4}$ of these SMSs would need to be visible to JWST to account for the observed population of Little Red Dot galaxies.
\end{abstract}


\maketitle
\section{Introduction} 
\label{sec:intro}
\noindent 
Ongoing debate regarding the origin of supermassive black holes (> 10$^6$ M$_\odot$) observed at early redshifts $z \sim 10$ \citep{Maiolino2024, Juodzbalis2024, Kovacs2024, Bogdan2024, Taylor2025, Ortiz2025} is split between the rapid growth of black hole remnants from the first Population III stars \citep{Madau2001, Smith2018, Banik2019, Singh2023, Mehta2024, Cammelli2025, Mehta2026} up from their initial masses of a few tens of M$_\odot$ \citep{Wollenberg2020, Jaura2022, Prole2022, Prole2022a, Prole2023}, or from theoretical `heavy seeds' with higher initial masses (> 10$^3$ M$_\odot$: \citealt{Regan2014, Wise2019, Latif2022, Prole2024a}). Investigations into the latter suggest multiple formation mechanisms for such objects, notably mergers in dense stellar clusters \citep{Antonini2019, Fragione2022, Schleicher2023, Rantala2025a, Paiella2026}, quasi-stars \citep{Begelman2010, Coughlin2024, Begelman2026, Hassan2026}, or rapidly accreting supermassive stars (SMSs:  \citealt{Shibata2002, Hosokawa2013, Umeda2016, Woods2017, Haemmerle2021}).

Recent observations by JWST have revealed a wealth of black hole host galaxies at high redshift (z $\gtrsim 8$). Typically, candidate galaxies are first identified through their photometric properties (e.g. UV-selected galaxies at high redshift), with spectroscopic follow-up subsequently used to identify broad H-$\alpha$ emission. The width and luminosity of the broad H-$\alpha$ line have been shown to provide an excellent quantitative estimator of black hole mass, at least at low redshift \citep{Greene2005}. However, selection simply on galaxy "colour" is also possible and indeed (where it is possible) much more efficient. The discovery of a new category of compact galaxies known as `Little Red Dots' (LRDs: \citealt{Matthee2024, Labbe2023,Akins2024,Guia2024, Perez-Gonzalez2024, Volonteri2024,Ananna2024,Baggen2024,Kocevski2025,Taylor2025a,Taylor2025}) has potentially aided the fast selection of AGN host galaxies (at least for this subset). LRDs are characterised by their spectral features which include a characteristic V shape, a sharp Balmer break, broad Balmer emission and non-detections in the X-ray bands - combined with their compact morphology (i.e. unresolved emission in the rest frame optical) and their red colour. Their red colours make them relatively easy to identify photometrically. However, their combination of features as noted above has proved difficult to explain with standard templates, and various models yield many interpretations of their spectra. Leading explanations include an AGN with complex dust extinction \citep{Inayoshi2025a,Inayoshi2025, Naidu2025, DeGraaff2025, DeGraff2025a, Rusakov2025, Rusakov2026, Sun2026}, extreme starburst galaxies \citep{Baggen2024, Guia2024}, and, more recently, quasi-stars \citep{Begelman2026, Hassan2026}, as well as main sequence SMS models \citep{Nandal2026, Nandal2026c, Zwick2026}. If LRDs are indeed ultimately powered by central AGN activity (as the evidence currently suggests) then they potentially give us a new observational window within which to categorise and probe massive black hole formation.\\
\indent The heavy seed formation model employed in \texttt{SEEDZ} allows us to investigate multiple possible 
origins for LRDs and indeed other high redshift UV and optically selected sources. As already noted, LRDs may arise from rapid accretion onto a massive black hole at the galactic centre, SMS activity, or perhaps a combination of these (linked) processes. The existence of SMSs (or quasi-stars) would not only provide a formation pathway for the observed SMBHs at high redshift but may also explain the electromagnetic spectra from LRDs.  \\
\indent  Accretion rates in excess of $\dot{\rm M}\gtrsim0.02$  M$_\odot$ yr$^{-1}$ onto a growing SMS will cause the stellar envelope to remain inflated with a cool photosphere and weak ionising feedback \citep{Hosokawa2013, Umeda2016, Woods2017, Haemmerle2021, Nandal2023}. This rapid accretion allows continued growth toward $10^{4}$--$10^{5} {\rm M_\odot}$ before the general relativistic instability eventually triggers collapse. Although metals promote fragmentation, recent simulations show that metal enriched clouds can still form SMSs through super-competitive accretion up to $Z\lesssim10^{-3}Z_\odot$ \citep{Chon2025}. Stellar evolution models extend this picture by showing that metal-enriched SMSs can remain cool and weakly ionising up to $Z\simeq10^{-2}Z_\odot$ \citep{Nandal2026a}. The required inflow of $\dot{\rm M}\gtrsim0.02$  M$_\odot$ yr$^{-1}$ is demanding, but it is not exceptional as shown in simulations of early galaxy assembly \citep[e.g.][]{Regan2014, Wise2019, Latif2022, Prole2024a, Prole2024b, Prole2026}.  \\
\indent Following the SMS phase, the star is expected to directly collapse into a massive black hole, which can continue to accrete surrounding material. Depending on the 
environmental conditions surrounding the newly formed massive black hole, the emission spectrum can continue to produce an LRD-like spectrum, or the accretion feedback could trigger the transition of the host galaxy from an LRD-like state to a more standard Optical/UV selected AGN, possibly via an intermediate stage of what has been coined a `little-blue-dot' (LBD: \citealt{Brazzini2026}).  Both the progenitor phase through the SMS pathway and the subsequent accretion onto the SMBH have been shown to produce spectral characteristics consistent with current LRD observations. \\
\indent While inflow rates are undoubtedly required for SMS star formation, it is less clear how efficient the formation of a massive central star is in practice. Metal enrichment will induce fragmentation and curb the efficiency of single object formation. As the metal enrichment of any collapsing gas cloud increases, fragmentation may become sufficiently widespread to lead to the formation of a dense stellar cluster. Depending on the density of the stellar cluster and the total number of stars, this configuration may also lead to the formation of a massive black hole at the cluster centre, which can grow via stellar capture \citep[e.g.][]{Rantala2025a, Rantala2026a, Paiella2026}. If, after formation, the initially relatively light black hole can accrete gas rapidly, its spectrum could produce the necessary LRD features without the need for a SMS progenitor. In this case we get a rapidly accreting black hole at the centre of the host galaxy, which also has a significant stellar component. The combination of an accreting black hole and a rich stellar environment could also explain observed LRD spectra.  \\
\indent Differentiating the possible pathways to heavy seed formation is a challenging problem, as the formation process itself occurs at small scales and at high redshift, making it generally inaccessible to current observations \citep[e.g.][]{Regan2024a}. Modelling various pathways numerically allows us to test models for consistency with observations. The \texttt{SEEDZ} model, presented here, is agnostic to the underlying seed formation pathway which occurs several orders of magnitude below the resolution of our grid. Even trying to apply subgrid modelling to this process is intractable at the current time. Instead our model (based on high inflow rates, see \S \ref{sec:method}) picks up the growth of the heavy seed seed following a possible progenitor phase involving SMS and/or dense stellar cluster formation. In this study, we aim to find out what conditions and halo assembly histories facilitate these high inflow rates in \texttt{SEEDZ}. Due to the similarity between the numerical formation criteria of heavy seeds in \texttt{SEEDZ} and the physically motivated criteria for SMS formation from stellar models \citep{Hosokawa2013, Umeda2016, Woods2017, Haemmerle2021, Nandal2023}, we can also use the \texttt{SEEDZ} simulation framework to identify potential SMS and dense stellar cluster formation sites across cosmic history and compare the environmental conditions required to the conditions observed around high redshift LRDs. We note however that at our resolution distinguishing between SMS formation and dense stellar cluster formation is likely to be challenging. 

\begin{table*}[]
\centering
\caption{Comparison between the Calibration and Highres simulation suits. From left to right - simulation suite name, minimum gas cell length, gas gravitational softening length, dark matter mass resolution and DM gravitational softening. All length are given as their co-moving values, at $z=10$ these lengths are a factor of 11 smaller.}
\begin{tabular}{ c c c c c c }
 Suite & V$_{\rm zoom}$ [cMpc$^{3}$] & $\Delta$x$_{\rm gas}$ [cpc] & S$_{\rm gas}$ [cpc] & M$_{\rm DM}$ [M$_\odot$] & S$_{\rm DM}$ [cpc]\\ 
 \hline
 \\
 \texttt{Calibration} & 705  & 91.7  & 222.6  & $1.2 \times 10^5$  & 4599.4 \\  
 \\
 \texttt{HighRes} & 10  & 7.8  & 19.3  & $1.5 \times 10^4$  & 192.9 
 \\
\label{table:1}
\end{tabular}
\end{table*}

\section{Numerical method}
\label{sec:method}
\noindent The full methodology for the \texttt{SEEDZ} simulation framework was presented in \cite{Prole2026} (hereafter \citetalias{Prole2026}) for the \texttt{Calibration} simulations. In this work, we present results based on the \texttt{HighRes} simulation suite. The \texttt{HighRes} simulations suite improves the dark matter particle mass resolution and hydrodynamic gas cell resolution by almost an order of magnitude compared to the initial \texttt{Calibration} simulations. Using the \texttt{HighRes} setup we can more accurately resolve the flow of gas into collapsing halos compared to \citetalias{Prole2026}. A comparison of the simulations parameters between the \texttt{Calibration} and the \texttt{HighRes} is given in Table \ref{table:1}.

In the dataset analysed here, we do not include black hole feedback to keep the physical interpretation of our `heavy seed' particles open to include SMSs or dense stellar clusters within the first few Myr after their formation, where no/very little feedback is expected. As presented in \citet{Prole2026a}, AGN feedback is also particularly destructive to the environments surrounding newly formed, growing heavy seeds. Analysing the datasets in which this feature is disabled also allows us to examine formation environments which remain largely unchanged during the first few Myr after formation, so that their properties can be analysed in that context. Additional datasets featuring different AGN prescriptions are also available and will be presented in a forthcoming analysis (Metha et al. in prep).

\begin{figure*}
  \centering

  \hbox{\hspace{0.2cm}\includegraphics[width=\linewidth]{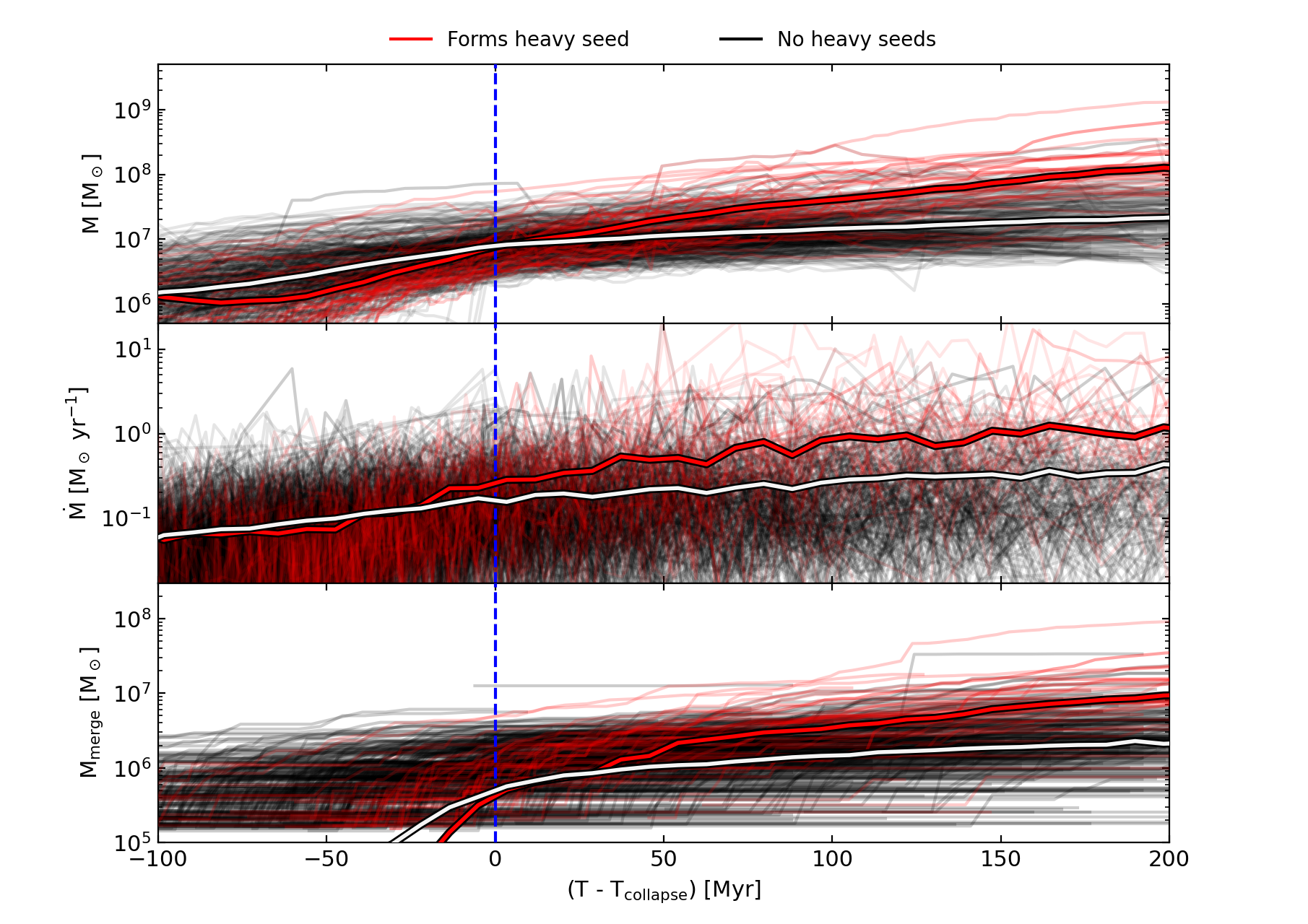}}
 
\caption[]
{\label{fig:heavy} Halo assembly histories taken from the merger tree. All quantities are plotted as a function of time relative to the time at which the halo formed its first star, T$_{\rm collapse}$. Lines are colour-coded to indicate whether the halo forms a heavy seed at any point during the simulation (red) or never forms a heavy seed (black). The three panels show the evolution of the halo mass (top), the halo growth rate (middle) and the cumulative mass gained from mergers (bottom). For each colour category, we show individual halos as transparent lines, and the average as a thick bold line (note the average lines for non-heavy seed-forming halos are white for better visibility). \\ \\ }
\end{figure*}


\begin{figure}
  \centering

  \includegraphics[width=\linewidth]{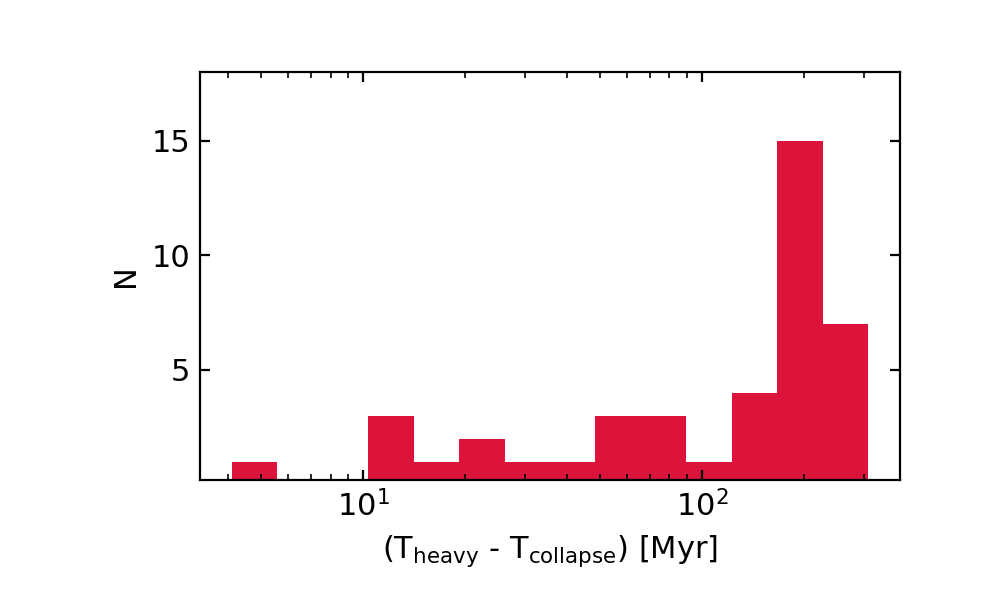}
 
\caption[]
{\label{fig:times} Histogram of the time elapsed between a host halo's initial collapse and the formation of its first heavy seed. \\ }
\end{figure}

\begin{figure}
  \centering

  \includegraphics[width=0.9\linewidth]{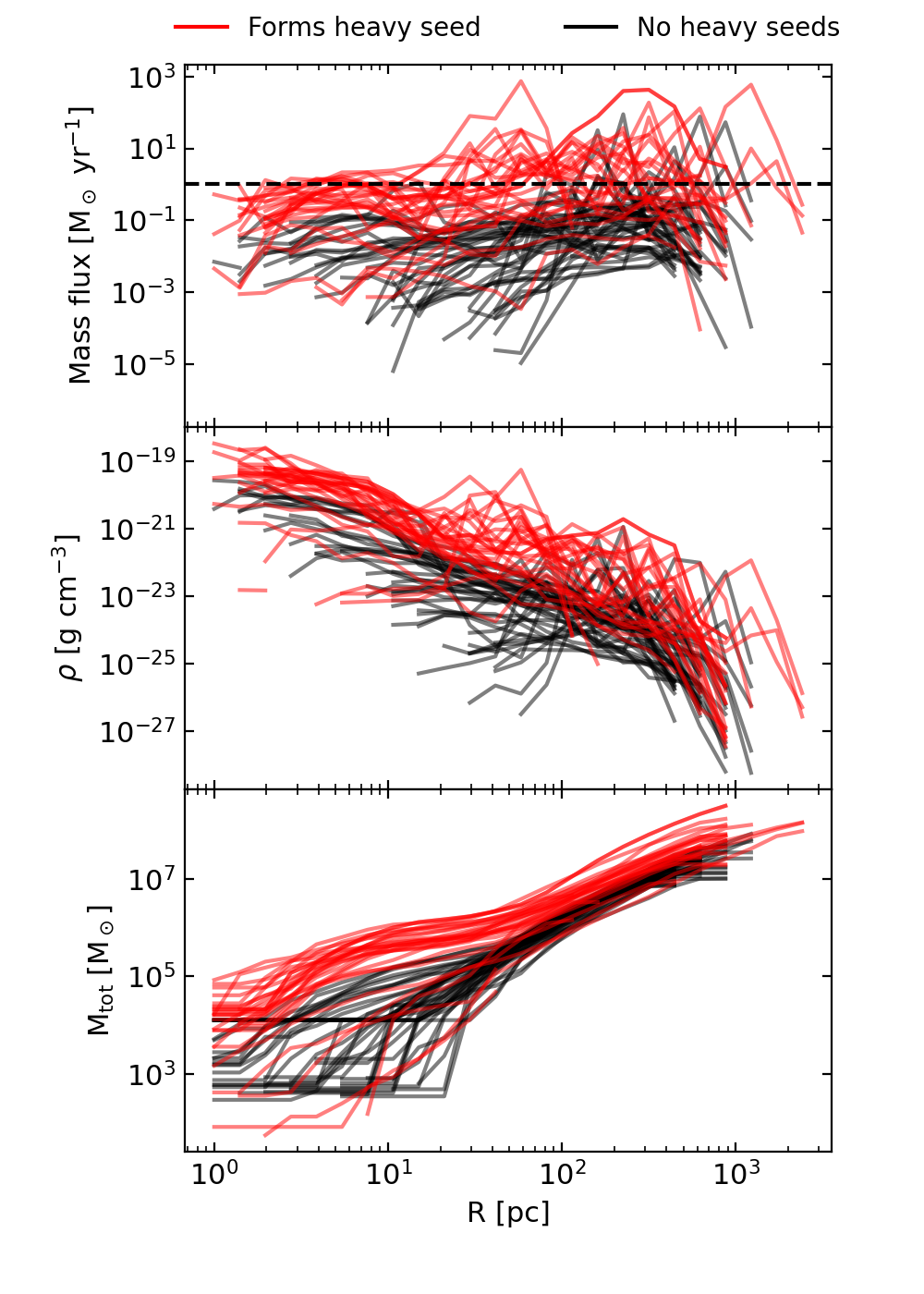}
 
\caption[]
{\label{fig:profile} Radial profiles of halos at the point of heavy seed formation (red) compared against non-heavy seed-forming halos sampled 100 Myr after they form their first star (black). We show the mass inflow rate (top), the density (middle) and the cumulative mass (bottom) profiles for each halo. \\ \\ }
\end{figure}

\begin{figure}
  \centering

  \includegraphics[width=0.9\linewidth]{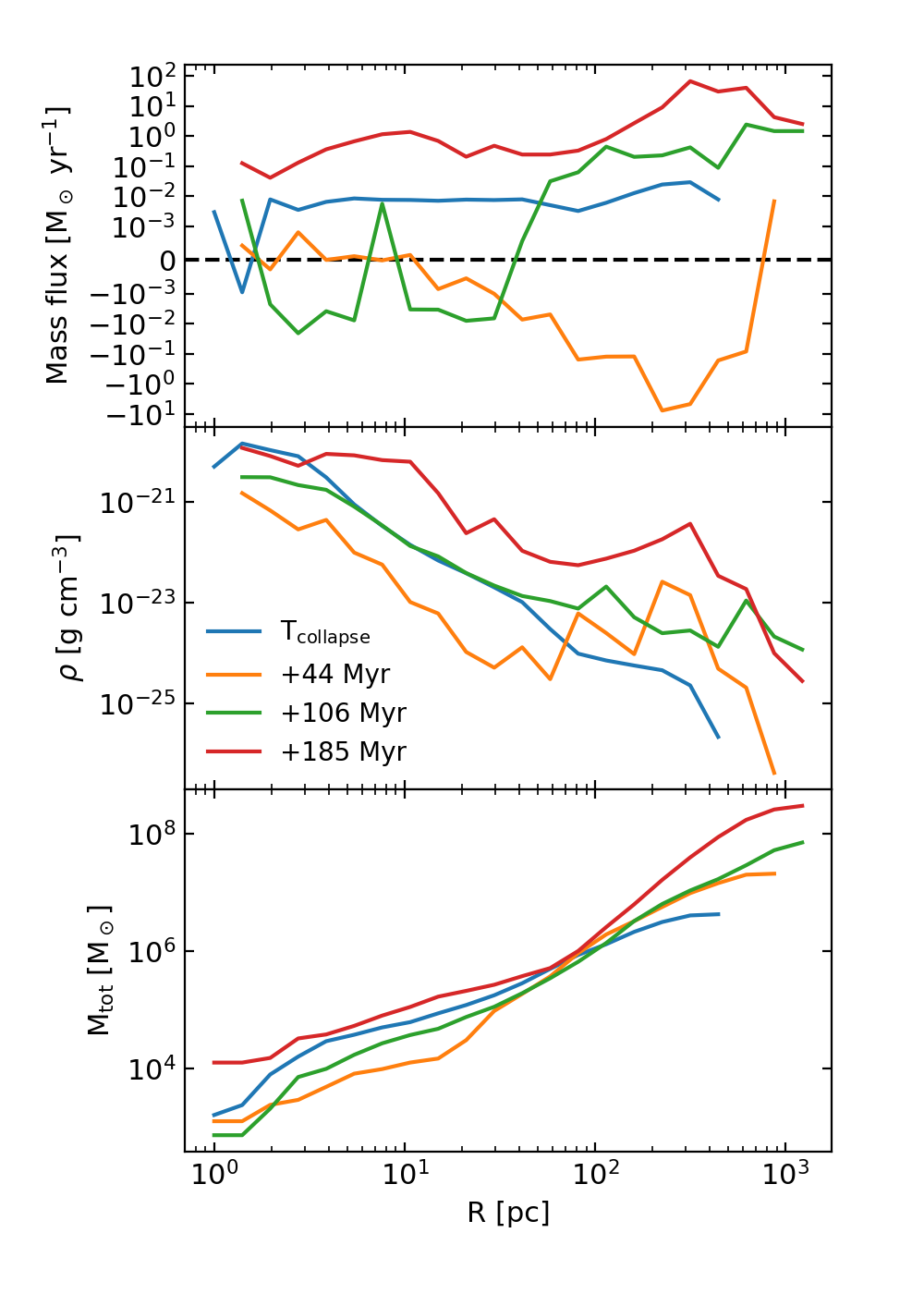}
 
\caption[]
{\label{fig:profile_time} Radial profiles for a single halo at different times between forming its first star (blue) to forming its first heavy seed (red). We show the mass inflow rate (top), the density (middle) and the cumulative mass (bottom) profiles at each time. \\ \\ }
\end{figure}


 

\begin{figure*}
  \centering

  \includegraphics[width=0.85\linewidth]{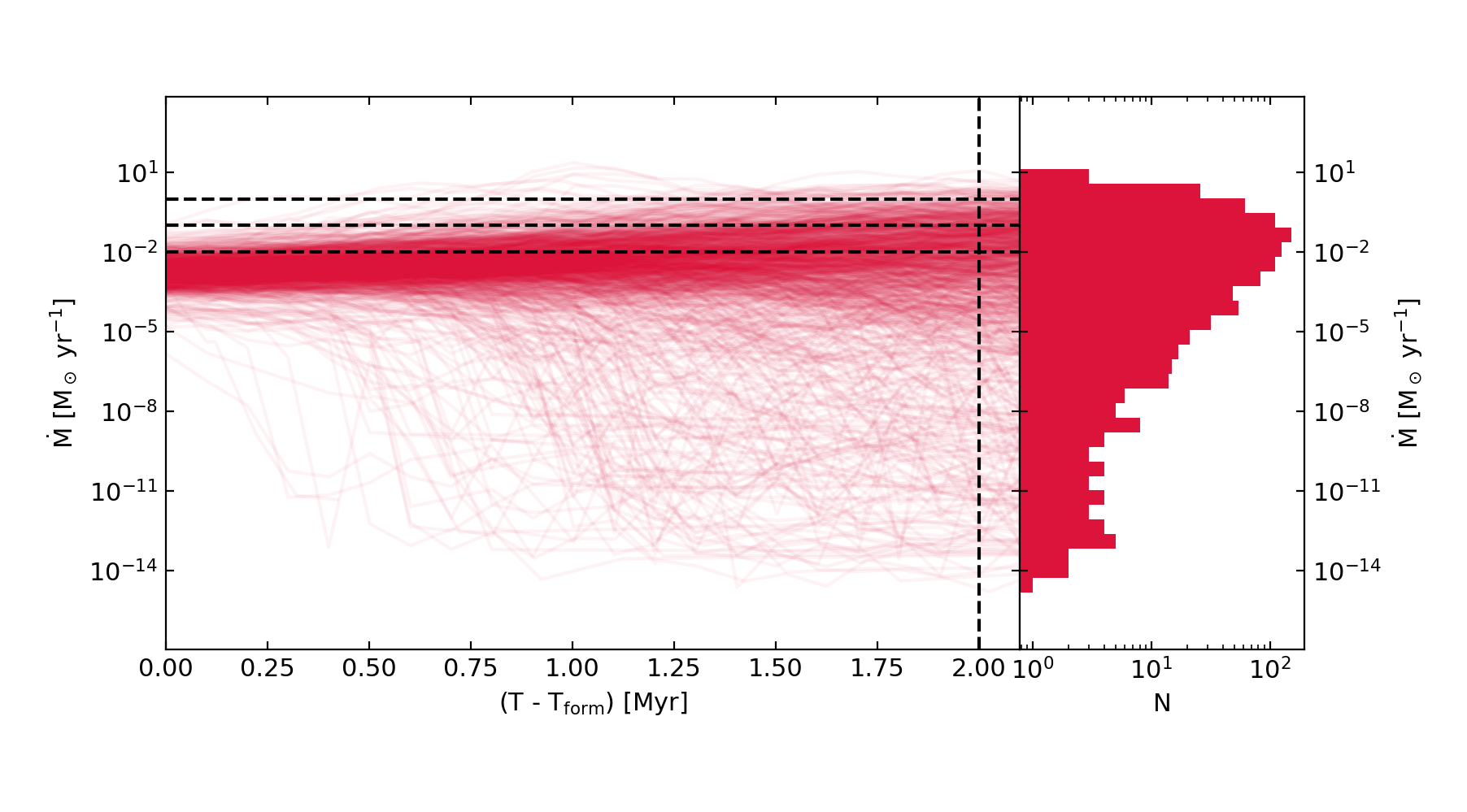}
 
\caption[]
{\label{fig:accretion} Initial growth rates of heavy seeds. Left: the accretion rate onto heavy seeds within the assumed 2 Myr lifetime of a SMS after its formation. Right: histogram of the accretion rates onto heavy seeds at 2 Myr after their formation. }
\end{figure*}


\begin{figure*}
  \centering

  \hbox{\hspace{1cm}\includegraphics[width=0.9\linewidth]{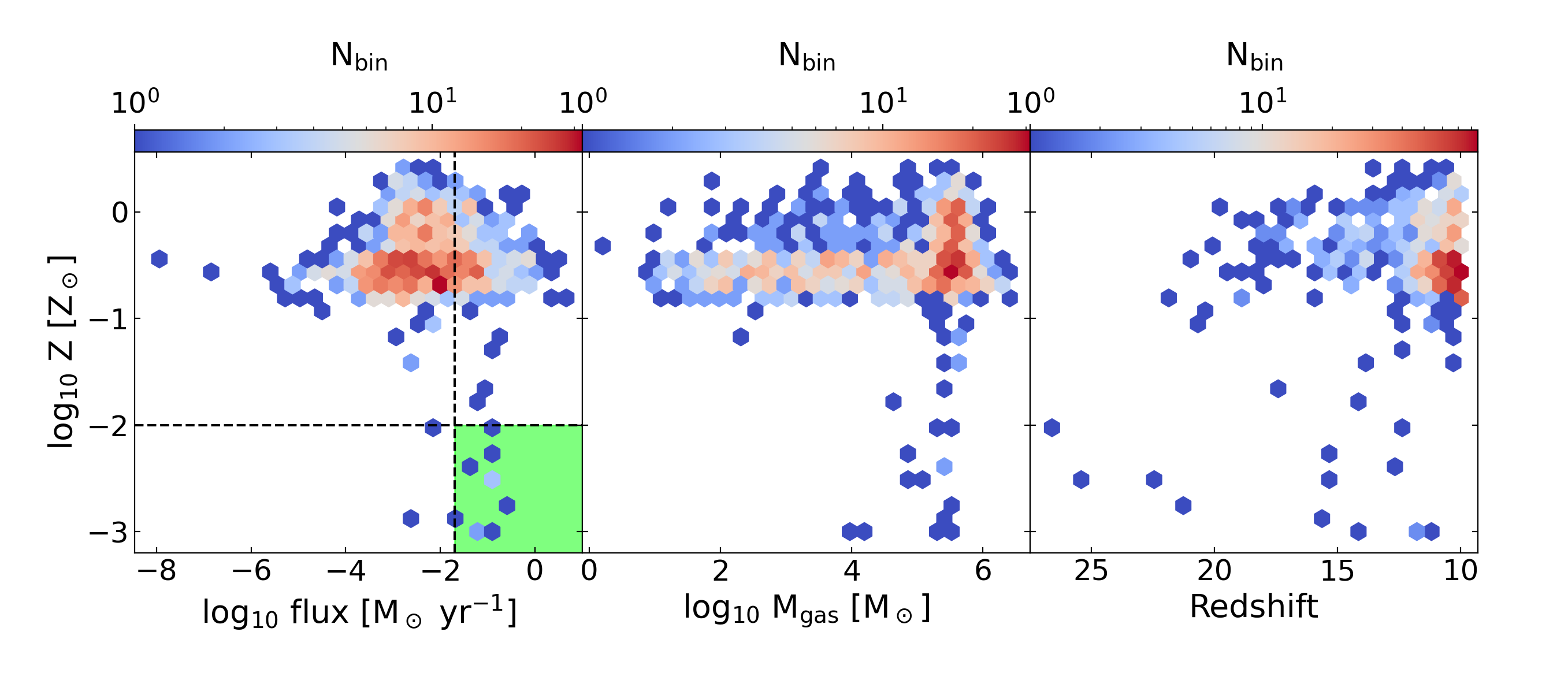}}
 
\caption[]
{\label{fig:metal} 2D heatmaps for heavy seed formation metallcities taken from the host gas cell at the time of formation. Left - average heavy seed growth rates within the first 2 Myr after formation. Middle - Total gas mass surrounding the heavy seed within 10 pc. Right - Redshift at the time of formation. We show the region of the 2D space where SMS is possible as a green square.
\\ \\ }
\end{figure*}


\begin{figure*}[]
\centering

  \includegraphics[width=1\linewidth]{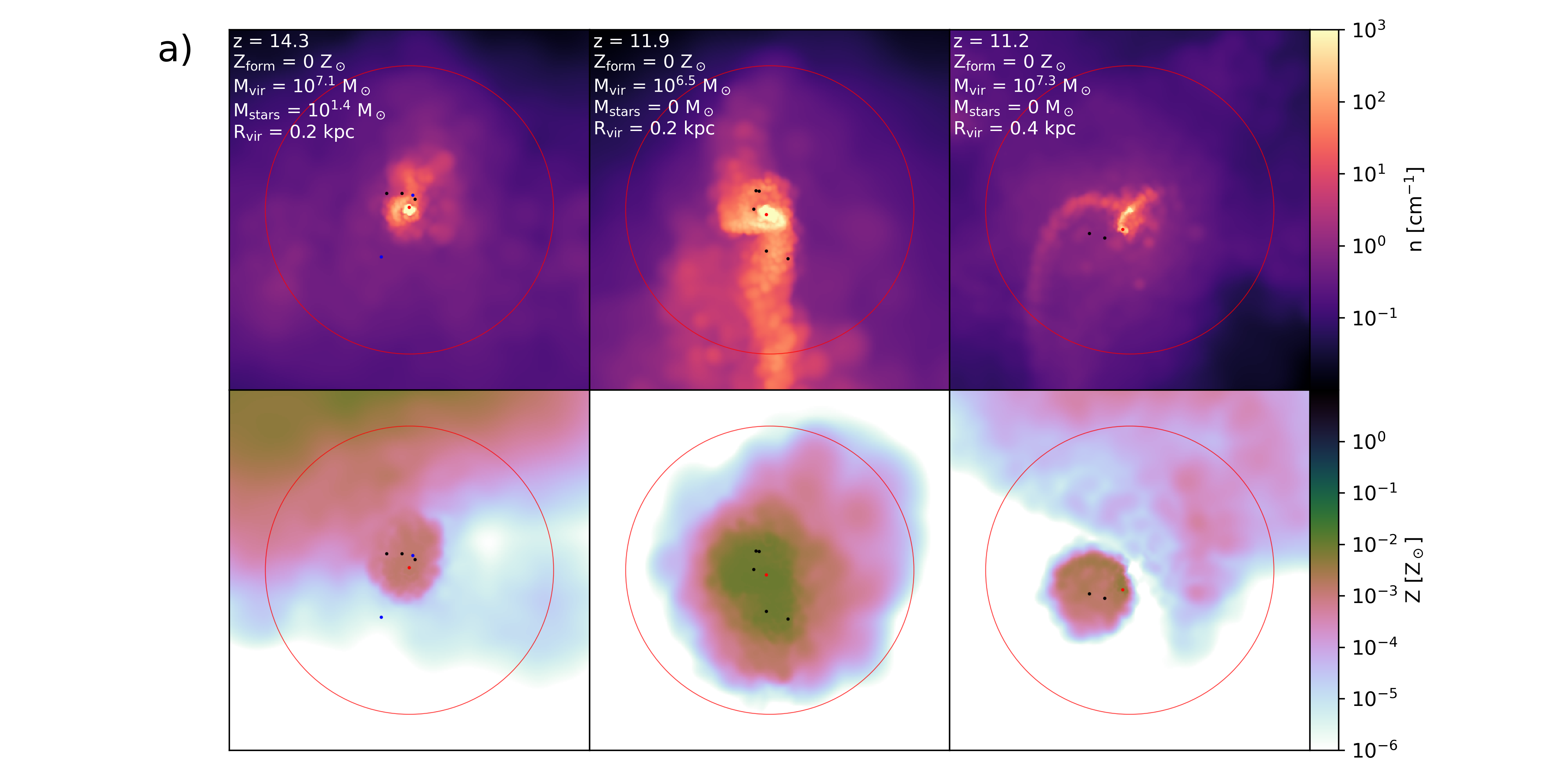}

  \includegraphics[width=1\linewidth]{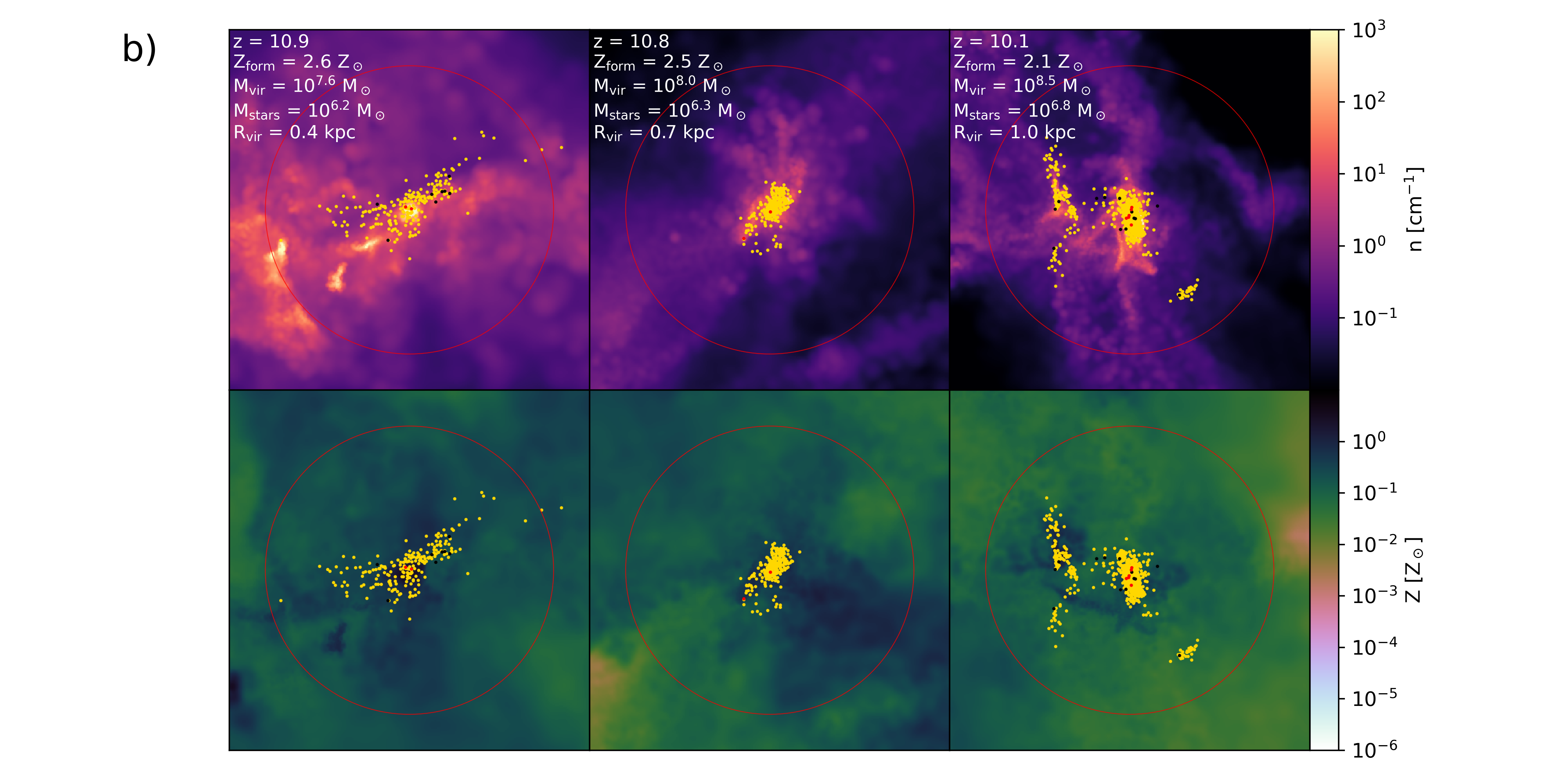}

\caption[]{\label{fig:projection} density (upper) and metallicity (lower) projections of heavy seed formation sites, taken from the next snapshot after their formation. a) three examples of heavy seeds forming in metal-free gas. b) three examples of heavy seeds forming in super-solar metallicity gas. Red dots represent heavy seeds, while black dots show light seed black holes, orange dots show PopII cluster particles, and blue dots show PopIII stars. We give the redshift, formation metallicity, total mass within the virial radius, stellar mass and the virial radius as white text. The virial radius of the halo is shown as a red circle. \\ }

\end{figure*}

\begin{figure}
  \centering

  \includegraphics[width=0.95\linewidth]{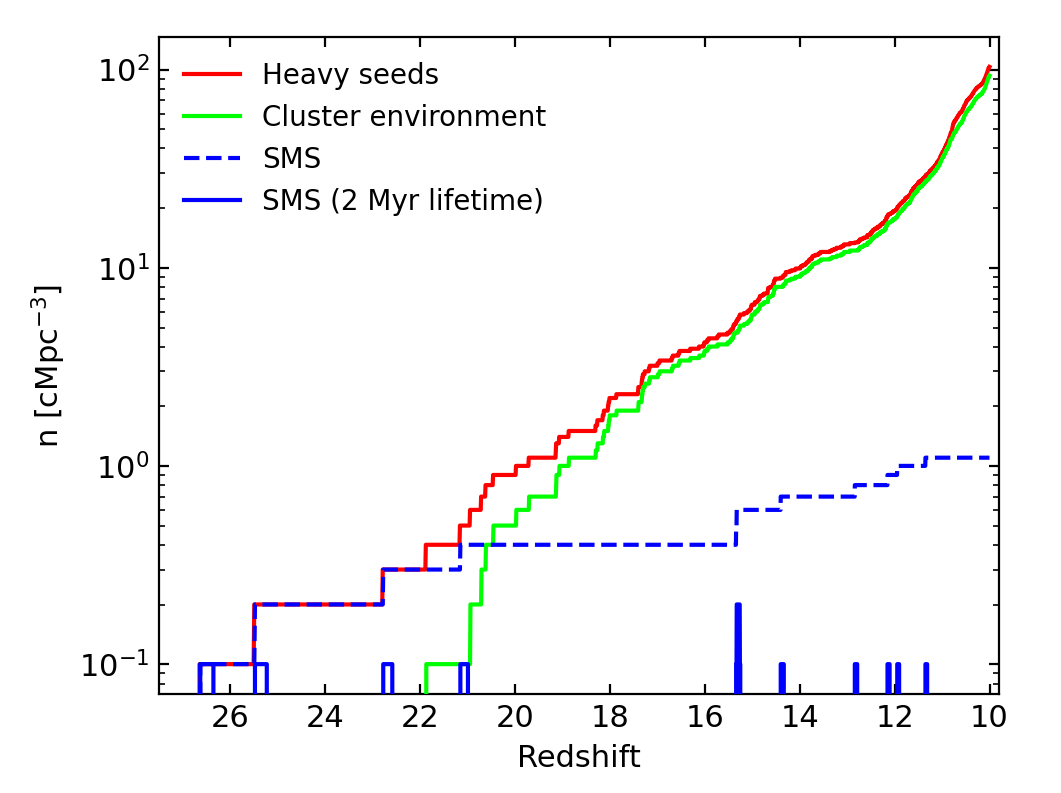}
 
\caption[]
{\label{fig:numberdensity} Number densities of all heavy seeds (red), those that form in regions of metallicity above 10$^{-2}$ Z$_\odot$ that are assumed to form from dense stellar clusters (green), those that meet the criteria for SMS formation i.e. metallicity below 10$^{-2}$ Z$_\odot$ and sustained inflow rates of 0.02 M$_\odot$ yr$^{-1}$ within the first 2 Myr life of a SMS (blue dashed), and the SMS number density assuming a stellar lifetime of 2 Myr (blue solid).\\ \\ }
 
\end{figure}

\subsection{Global parameters}
\noindent As already discussed the simulation dataset analysed here is taken from the \texttt{HighRes} \texttt{SEEDZ} simulations to be presented more fully in Metha et al. in prep. The same parent simulation box from \citetalias{Prole2026} was used, which has a side length of 40 cMpc h$^{-1}$. We focus on one of the original zoom-in locations from the calibration runs from \citetalias{Prole2026}, denoted as the \texttt{Normal1} region. From this, a high-resolution zoom-in simulation was initialised within a Lagrangian region of dimensions 1.25 $×$ 2.02 $×$ 1.21 (cMpc)$^{3}$ h$^{−3}$ at $z = 127$ (see Table \ref{table:1}). The region was then regenerated at higher resolution using MUSIC \citep{Hahn2011}. The simulations were performed with the developer version of \texttt{AREPO2} \citep{Springel2010, Pakmor2016,Pakmor2023}. In the high resolution region, dark matter particles have masses of $1.0 \times 10^4$ M$_\odot$ h$^{-1}$ and a gravitational softening length of 130 cpc h$^{-1}$ (approximately 11 pc physical at z = 10), roughly a factor of 10 better than the original \texttt{Calibration} suite of \texttt{SEEDZ} simulations. During the simulation, the gas cell mesh was refined such that the Jeans length was resolved by at least 4 cells down to the minimum cell size, which was set to 5 pc h$^{-1}$ (0.67 pc at $z=10$), with a gas gravitational softening length of 13 cpc h$^{-1}$. The simulations were run to $z=10$.

\subsection{Stars and black holes}
\noindent As regions of gas become denser under gravitational collapse, the simulation mesh refines locally. Gas cells that become gravitationally unstable after reaching the maximum level of refinement trigger the formation of a sink particle. Depending on the environmental conditions at collapse, the sink particle type is determined as follows:

   \begin{itemize}
      \item  If the radial mass flux within a 10~pc sphere surrounding the candidate cell exceeds 1 M$_\odot$ yr$^{-1}$, a heavy seed black hole is formed (with the progenitor unresolved, but possible interpretations include a SMS or dense stellar cluster, depending on the fragmentation processes below the resolution level of our simulations). The object's mass is sampled from a power law with a -0.5 exponent between 5 $\times$ 10$^3$ M$_\odot$ and 10$^5$ M$_\odot$. A sink particle is formed and the appropriate mass is taken from the host gas cell.
      \item When the radial flux is less than 1 M$_\odot$ yr$^{-1}$, and the metallicity within the 10~pc sphere is less than a threshold of 10$^{-4}$ Z$_\odot$, an individual Population III (PopIII) star is formed. Its mass is sampled from a top-heavy M$^{-1.3}$ IMF between 1 - 300 M$_\odot$.
      \item When the radial flux is lower than 1 M$_\odot$ yr$^{-1}$ but the metallicity is above 10$^{-4}$ Z$_\odot$, the cell is converted into a Population II (PopII) cluster particle, sampling stellar mass from a Kroupa IMF until the cluster mass is accounted for.
   \end{itemize}

In the case of PopII stars when the stellar mass is sampled, the stellar lifetime is calculated. When a star's age exceeds its lifetime:
    \begin{itemize}
    \item Stars with masses in the type II (11 - 20 M$_\odot$) supernovae (SNe) or the hypernova (20 - 40 M$_\odot$) mass bands will experience an explosion and undergo some degree of mass loss. The explosion energy and resulting black hole remnant mass are calculated from the stellar mass as taken from \cite{Nomoto2006}, and the explosion is modelled with an isotropic injection of thermal energy into the surrounding 10~pc. The particle is then converted into a black hole particle.
    \item Stars with masses in the pair instability mass band (140 - 260 M$_\odot$) have explosions so energetic that the remnant is destroyed, i.e. the particle is deleted after the momentum injection.
    \item Stars in the direct collapse mass band (40 - 140 M$_\odot$) collapse into black hole particles without experiencing an SNe or mass loss.
    \end{itemize}

Note that conversion of stellar particles into black holes only occurs for PopIII particles, not PopII clusters. Unlike stellar particles, black holes (PopIII remnants and heavy seeds) are able to accrete gas from their surroundings. Black hole particles have a Bondi-Hoyle accretion prescription with a vorticity adjustment, acting on gas cells within an accretion radius of 10~pc. For more details, see \citetalias{Prole2026}.

\section{Formation environments}
\label{sec:where}
\noindent Here we investigate the typical environmental conditions that lead to the high (1 M$_\odot$ yr$^{-1}$) inflow rates within 10 pc regions surrounding gas gravitational potential peaks, as required for our heavy seed black hole formation. We analyze the assembly history of the top 1000 most massive halos with merger trees using the halo finder \texttt{Subfind} \citep{Springel2021}. In Figure \ref{fig:heavy}, we show the growth of halos as a function of the time since they formed their first sink particle, T$_{\rm collapse}$. We separate the data into two groups, halos that form heavy seeds at any point in the simulation (red) and those halos that never form a heavy seed (black). We plot the average time evolution for both groups as a solid thick line.

From the sample of 1000 halos, 42 were found to form at least one heavy seed. On average, both groups form their first stars at the same halo mass of a few times 10$^6$ M$_\odot$, i.e.\ in the minihalo regime. However, in the following 100 Myr, the halos that go on to form heavy seeds grow at accelerated rates. Figure \ref{fig:times} shows how much time passes between the initial collapse (i.e.\ the formation of the first star) and the formation of the first heavy seed in each halo. Most heavy seeds form approximately 200 Myr after the first star, after the halos have distinguished themselves from the non-heavy seed-forming group. The bottom panel of Figure \ref{fig:heavy} shows the mass added to the main progenitor halo from mergers with subhalos, showing that heavy seed-forming halos also gain more mass via mergers.

Figure \ref{fig:profile} shows radial profiles for the most massive 100 halos by redshift 10, containing 42 heavy seed-forming halos and 58 non-heavy seed forming halos. For heavy seed-forming halos, we plot the profile at the point where they formed their first heavy seed, by tracing them back via the merger tree. For the non-heavy seed-forming group, we instead plot the profile 200 Myr after T$_{\rm collapse}$, as this was the most common timescale for heavy seed formation seen in Figure \ref{fig:times}. Due to the higher growth rates, heavy seed-forming halos experience higher mass inflow rates cascading in from their virial radii of a few 100~pc down to the central 1~pc, leading to higher central densities and the ability to form heavy seeds.

We show the time evolution of these radial profiles for the largest heavy seed-forming halo in Figure \ref{fig:profile_time}. At the time of initial collapse, the halo had modest inflow rates of 10$^{-2}$ M$_\odot$ yr$^{-1}$. The initial burst of star formation and subsequent supernovae explosions opposed inflow, and drove gas outwards, reducing the central densities. On timescales of 100 Myr after T$_{\rm collapse}$, these outflows begin to reverse back inwards at the outskirts of the halo (>100 pc), eventually exceeding inflow rates of 1 M$_\odot$ yr$^{-1}$ within the central 10~pc by the time of heavy seed formation at 185 Myr.


\section{Supermassive Stars \& Dense Stellar Clusters}
\label{sec:where}

\noindent We can consider heavy seeds formed in the \texttt{SEEDZ} simulations to form via a SMS or formation through collisions in a dense stellar cluster.  In reality, the difference between the two regimes likely depends on the metallicity of the formation site and the sustained inflow rate onto the central region. \\
\indent Typically sustained inflow rates of 0.02 M$_\odot$ yr$^{-1}$ are required for SMS formation \citep{Hosokawa2013, Umeda2016, Woods2017, Haemmerle2018, Nandal2023}, and viable SMS models now extend up to metallicities of 10$^{-2}$ Z$_\odot$ \citep{Chon2025, Nandal2026a}. Recently, \cite{Nandal2026a} showed that SMSs forming in increasing metallicities decrease the final stellar mass. At metallicities in excess of  10$^{-2}$ Z$_\odot$ the final SMS is reduced to approximately 2000 M$_\odot$ in their models. At even higher metallicities, the collapsing gas cloud is likely to undergo significant fragmentation, which will likely lead to the formation of a dense stellar cluster. Such environments have been shown to harbour massive black hole formation at similar masses ($10^3 - 10^4$ \msolarc) up to metallicities as high as 0.5 \zsolar \citep[e.g.][]{Rantala2026a}. The boundary where SMS formation becomes non-viable and instead a dense stellar cluster forms is not yet clear, but here we use the conditions that SMS formation requires metallicity below 10$^{-2}$ Z$_\odot$ and a mass inflow above 0.02 M$_\odot$ yr$^{-1}$.

The typical lifetime of a SMS is appproximately $\sim$ 2 Myr \citep{Yungelson2008, Yusof2013, Woods2020}. Figure \ref{fig:accretion} shows the growth rates of all heavy seeds within this initial 2 Myr period. At 2 Myr, most of the heavy seeds formed in our simulation have accretion rates of 0.1 M$_\odot$ yr$^{-1}$ with the distribution extending up to 1 M$_\odot$ yr$^{-1}$. We take the average growth rate within this period as the canonical mass flux in the SMS approximations performed below. While these growth rates are calculated within a 10 pc region, the real radius of SMSs is at maximum 200--300 AU \citep{Hosokawa2013, Nandal2026b} -- scales that we do not resolve in our simulations. Although the mass cascade from 10~pc scales down to the SMS is unresolved, the required efficiency can be estimated as
\begin{equation}
\epsilon_{\rm cas}
=
\frac{\dot{M}_{\star}}{\dot{M}_{10{\rm pc}}} .
\label{eq:epscas_def}
\end{equation}
Around 10$\%$ of the mass flux at 10 pc will need to make it's way to the stellar surface of the accreting SMS. Whether this can occur in practice or not will ultimately determine whether SMS star formation can occur.

We show the metallicity of the host gas cells at the time of heavy seed formation in Figure \ref{fig:metal}, and plot them against the average growth rates, surrounding gas mass within 10 pc, and formation redshift. We set the metallicity floor to $10^{-3}$ \zsolar for display purposes, as this is an order of magnitude below the threshold for SMS formation that we set here. From the 932 heavy seeds that form throughout the simulation, there are 13 that form in regions with metallicities below $10^{-2}$ \zsolar and 3 heavy seed formation sites with metallicity below $10^{-6}$ (pristine). We colour the region of the growth-metallicity space that permits SMS formation in green, containing only 11 SMS candidates. The bulk of heavy seeds form in environments with metallicities between 0.1 to a few times solar metallicity, while some rare heavy seeds continue to form in lower metallicity (< 10$^{-2}$ \zsolar) environments up until $z=10$. As heavy seeds typically form a few hundred Myr after the first stars in each halo (see Figure \ref{fig:times}), it is unsurprising that most heavy seeds formed within metal-enriched regions, even as early as $z \sim 25$, as stars with masses greater than $\sim$6 M$\odot$ have lifetimes shorter than 100 Myr, allowing supernovae explosions to enrich the halo prior to the formation of heavy seeds. These results taken at face value indicate that heavy seed black hole formation may well be dominated by collision induced formation, possibly through competitive accretion with only a minority driven by monolithic collapse and SMS formation.  

The middle panel shows the total gas mass contained within 10~pc surrounding the newly formed heavy seeds, acting as the reservoir from which the SMS or stellar cluster gains mass from. The most common occurrence is that heavy seeds form with 10$^5$-10$^6$ M$_\odot$ available to them within this region. This mass is sufficient to fuel either the formation of a SMS or stellar cluster.

We show density and metallicity projections of heavy seed formation sites with different metallicities in Figure \ref{fig:projection}. Panel a) shows three heavy seeds which form in <10$^{-6}$ Z$_\odot$ environments. These sites have no stellar mass components, but contain a few light seed black holes which have recently undergone SNe from their original PopIII states. Some of these stars were in the direct collapse mass band and hence did not add to the metal enrichment of the halo. These early galaxies are also either isolated from other systems and hence surrounded by pristine gas, or lie on the edge of a larger metal enriched region. In contrast, b) shows three galaxies forming heavy seeds in near-solar metallicity regions. These halos are heavily metal enriched on scales larger than their virial radii, and contain large stellar components in the range of 10$^6$-10$^7$ M$_\odot$. The formation of SMSs therefore relies on achieving high inflow rates while the galaxy is in its early stages, before a substantial population of stars builds up to enrich the halo.

\subsection{The Number Density of Supermassive Stars and/or Dense Stellar Clusters}
We now use the average growth rates and formation metallicities to calculate the number density of SMSs as a function of cosmic time, by only considering heavy seeds with average accretion rates above 0.02 M$_\odot$ yr$^{-1}$ within their first 2 Myr and a formation metallicity below $10^{-2}$ Z$_\odot$ to be SMS. We also assume that these objects exist only for typical SMS lifetimes of 2 Myr before transitioning into a massive black hole. We assume that for objects that do not fit this criteria, the black hole seed originates from collisions in a dense stellar cluster. \\
\indent Figure \ref{fig:numberdensity} compares the number density of all heavy seeds (red) against the density of dense stellar clusters (green) and SMSs (blue). Generally, only 1 SMS exists at a time in the simulation volume of 10 cMpc$^{3}$, giving a number density of 0.1 cMpc$^{-3}$. There are extended periods where no SMSs exist (due to their intrinsically short lifetimes). As the formation of SMSs is seemingly stochastic, we can assume that they would exist more consistently at this number density at all times if simulated in a much larger volume. By $z=10$, we find an overall heavy seed number density of 100 Mpc$^{-3}$, dominated almost entirely by high metallicity (>10$^{-2}$ Z$_\odot$) formation sites, that we assume lead to the formation of dense stellar clusters. 

Recently, the emission from SMSs has been suggested as a potential explanation for the unusual spectral features of LRD galaxies \citep{Nandal2026, Nandal2026c, Zwick2026}. Current observational estimates of the number density of LRD galaxies are of the order of $n_{\rm LRD} \sim 10^{-5}$ cMpc$^{-3}$ \citep{Schindler2025}. Although this value is valid at much lower redshifts ($z\sim 4-7$), directly comparing it our simulated SMS number densities implies that only a fraction of $\sim 10^{-4}$ of the SMSs would need to be visible to JWST in order to account for the observed LRD population.



\section{Conclusions}
\label{sec:conclusions}
Here we have presented a new cosmological simulation using the \texttt{SEEDZ} framework, with enhanced resolution compared to the original suite. The simulation was performed with the hydrodynamics code \texttt{Arepo2} and probes scales of $\sim$0.5 physical pc. On top of the base code, the \texttt{SEEDZ} framework includes subgrid models for PopIII star formation, supernovae explosions, metal enrichment, PopII star formation, black hole formation, and heavy seed black hole formation. Heavy seeds form in regions where the mass inflow into 10 pc regions surrounding a gravitationally unstable sub-pc sized cell exceeds 1 M$_\odot$ yr$^{-1}$. By splitting our catalogue of early galaxies into two groups -- those that form heavy seeds and those that do not -- we investigate the conditions necessary for high inflow environments to occur and by extension the conditions necessary for SMS and/or dense stellar cluster formation. We make the following conclusions:
    \begin{itemize}
    \item Halos that host heavy seed black holes grow more rapidly than their non-seed-forming counterparts in the few hundred Myr following initial collapse, quickly assembling total masses above $10^8$ \msolarc. Across all halos, star formation consistently predates the emergence of the first heavy seed.
    \item At the time of heavy seed formation, the mass inflow rate regularly exceeds 0.1 M$_\odot$ yr$^{-1}$ even into the inner 1 pc in most heavy seed-forming halos. Following the growth of heavy seeds for the first few Myr after their formation reveals that most are able to sustain accretion rates of 0.1 M$_\odot$ yr$^{-1}$.
    \item By assuming a SMS forms when a heavy seed forms in a region of metallicity below 10$^{-2}$ Z$_\odot$, and achieves an average accretion rate above 0.02 M$_\odot$ yr$^{-1}$ within the assumed initial lifetime of 2 Myr, we find a number density of SMSs of 0.1 cMpc$^{-3}$, compared to the overall number density of heavy seeds of 100 cMpc$^{-3}$. Therefore, metal-poor SMS formation is likely a small percentage of total heavy seed formation. 
    \item The bulk of our heavy seeds form in metal enriched (> 10$^{-2}$ Z$_\odot$) environments and are therefore expected to form from less massive heavy seeds in the range $10^{3} - 10^{4}$
    \msolar. Such a seed can be formed either through metal-enriched, massive star formation as modelled by \cite{Chon2025} or perhaps through collisions in dense stellar environments \citep[e.g.][]{Rantala2025b, Rantala2026a}.
    \item Throughout our simulations (even at very high redshift) heavy seed formation occurs in regions with metallicities in excess of $10^{-2}$ Z$_\odot$. Of the 932 heavy seeds that form
    throughout the simulation, only 13 form in regions with metallicities below $10^{-2}$ Z$_\odot$, and only 3 are truly metal-free.
    \end{itemize} 
    
\vspace{0.2cm}
\section*{Acknowledgements}
\noindent JR \& JB acknowledges support from the Royal Society and Research Ireland under grant number 
 URF\textbackslash R1\textbackslash 191132. LP, DM \& JR acknowledge support from the Research Ireland Laureate programme under grant number IRCLA/2022/1165. JHW acknowledges support from NSF grants AST-2108020 and AST-2510197 and NASA grant 80NSSC21K1053. RSB acknowledges support from  UKRI Future Leaders Fellowship MR/Y015517/1. DN was supported by the Swiss National Science Fund (SNSF) Postdoctoral Fellowship, grant number: P500-2235464.
 \ \
The simulations were performed on the Luxembourg national supercomputer MeluXina and the Czech Republic EuroHPC machine Karolina.
The authors gratefully acknowledge the LuxProvide teams for their expert support.
\ \ 
The authors wish to acknowledge the Irish Centre for High-End Computing (ICHEC) for the provision of computational facilities and support.
\ \
The authors acknowledge the EuroHPC Joint Undertaking for awarding this project access to the EuroHPC supercomputer Karolina, hosted by IT4Innovations through a EuroHPC Regular Access call (EHPC-REG-2023R03-103) and to the LuxProvide supercomputer Meluxina  through a EuroHPC Regular Access call (EHPC-REG-2025R01-008).
\ \
SK has been supported by a Research Fellowship from the Royal Commission for the Exhibition of 1851.
\ \ 
MAB is supported by a UKRI Stephen Hawking Fellowship (EP/X04257X/1).
\ \ 
DS acknowledges support from the Science and Technology Facilities Council (STFC) under grant ST/W000997/1.
\ \
RSK acknowledges financial support from the European Research Council via the ERC Synergy Grant ``ECOGAL'' (project ID 855130),  from the German Excellence Strategy via the Heidelberg Cluster of Excellence (EXC 2181 - 390900948) ``STRUCTURES'', from the German Science Foundation under grant KL 1358/22-1, and from the German Federal Ministry for Economic Affairs and Energy in project ``MAINN'' (funding ID 50OO2206). RSK is grateful for computing resources provided by the Ministry of Science, Research and the Arts (MWK) of the State of Baden-W\"{u}rttemberg through bwHPC and the German Science Foundation (DFG) through grants INST 35/1134-1 FUGG and 35/1597-1 FUGG, and also for data storage at SDS@hd funded through grants INST 35/1314-1 FUGG and INST 35/1503-1 FUGG. RSK also thanks for computing time provided by the Leibniz Rechenzentrum via grants pr32lo, pr73fi and GCS large-scale project 10391. 
\ \ 
PCC acknowledges support from a STFC Small Award (UKRI1187): ``Probing the origins of stars and life with a new approach to chemical modelling''.
\ \
MGH has been supported by STFC consolidated grants ST/N000927/1 and ST/S000623/1.
\vspace{0.5cm}

\bibliographystyle{aasjournal}
\bibliography{SEEDZ_References}



\end{document}